\documentclass{aa}
\usepackage[dvips]{graphicx}
\usepackage{txfonts}
\usepackage{longtable,lscape}
\usepackage{natbib}
\usepackage{cases}
\usepackage{subfig}
\bibpunct{(}{)}{;}{a}{}{,}
\usepackage{color}
\usepackage{url}
\def\be{\begin{equation}}
\def\ee{\end{equation}}

\begin{document}
   \title{Str\"{o}mgren photometric survey in the\\Galactic anticenter
direction 
\thanks{Final catalog and catalog with individual measurements are only available in electronic form
at the CDS via anonymous ftp to cdsarc.u-strasbg.fr (130.79.128.5)
or via http://cdsweb.u-strasbg.fr/cgi-bin/qcat?J/A+A/} \fnmsep
\thanks{Based on observations made with the Isaac Newton Telescope operated on
the island of La Palma by the Isaac Newton Group in the Spanish Observatorio del
Roque de los Muchachos of the Instituto de Astrof\'{\i}sica de Canarias.}
} 
 \author{M. Mongui\'{o} \inst{\ref{inst1}}
      \and F. Figueras \inst{\ref{inst1}}
      \and P. Grosb\o{}l \inst{\ref{inst2}}
          }
   \offprints{M. Mongui\'{o},
   \email{mmonguio@am.ub.es}}
   \institute{Departament d'Astronomia i Meteorologia and IEEC-ICC-UB,
     Universitat de Barcelona,
     Mart\'i i Franqu\`es, 1, E-08028 Barcelona, Spain \label{inst1} \and
 European Southern Observatory, Karl-Schwarzschild-Str. 2, D-85748 Garching,
Germany \label{inst2}}
   \date{Received  / Accepted }
   \abstract
{}
{The main purpose is to map the radial variation of the stellar space density 
for the young stellar population in the Galactic anticenter direction in 
order to understand the structure and location of the Perseus spiral arm.}
{A $uvbyH\beta$ Str\"{o}mgren photometric survey covering 16$\degr ^2$ in the
anticenter direction was carried out using the Wide Field Camera at the
Isaac Newton Telescope. This is the natural photometric system for 
identifying young stars and obtaining accurate estimates of individual distances and ages.
The calibration to the standard system was undertaken using open clusters.}
{We present a main catalog of 35974 stars with all Str\"{o}mgren indexes and a more extended one with
96980 stars with partial data. The central 8$\degr ^2$ have a limiting magnitude of V$\sim$17$^m$,
while the outer region reaches V$\sim$15$\fm$5. 
These large samples 
will permit us to analyze the stellar surface density variation associated to the Perseus arm 
also to study the
 properties of the stellar component  and the  interstellar extinction in the anticenter direction.
}
{}

\maketitle

\section{Introduction}\label{intro}
While it is well established that spiral arms are important agents
driving the evolution of the galactic disks \citep{2011MNRAS.410.1637S, 2011ApJ...730..109F}, the
observational evidence of the spiral arms in the Milky Way are frustratingly inconclusive 
\citep{2011MNRAS.417..698L}.
Key questions are still open: e.g.,
which is the mechanism of the formation and evolution of the spiral pattern in stellar disks; are they transients or
long-lived structures;
which are their building blocks, stellar or gaseous? Spitzer/IRAC infrared data 
\citep{2008ASPC..387..375B} provide new insight into the spiral pattern
in the inner  region of the Galactic disk,
but we have a vague and very confused picture of the outer Milky Way spiral arm
structure. 

Few projects have studied the anticenter, such as the VLBA project \citep{2009ApJ...700..137R, 2011AN....332..461B}
that provides accurate parallaxes for some masers in massive star-forming regions. 
\cite{2008ApJ...672..930V} looked at the third Galactic quadrant using kinematic distances of CO clouds
and some open clusters and associations, and propose that the Perseus spiral arm is only
defined by gas over a large extent in this direction. 
Maps of OB-associations and
HII-regions \citep{1976A&A....49...57G, 2003A&A...397..133R} and the Galactic distribution of free
electrons 
\citep{1993ApJ...411..674T}
show a four-armed pattern. 
On the other hand, COBE K-band data \citep{2000A&A...358L..13D} suggest that
the non-axisymmetric mass perturbation has a two- rather than a four-armed structure. External galaxies often show a
two-armed structure
in near-infrared, while they may appear multi-armed in visual bands
\citep{2004A&A...423..849G}.
The spiral
model of the Milky Way obtained with Spitzer/IRAC infrared data in the Galactic
center
direction \citep{2008ASPC..387..375B} agrees with this extragalactic
scenario.
\cite{2008ASPC..387..375B} proposes that the Milky Way has two
major spiral arms (Scutum-Centaurus and Perseus) with higher stellar
densities and two minor
arms (Sagittarius and Norma) mainly filled with gas and pockets of young stars.

The Str\"{o}mgren photometric survey presented here was conducted with the main goal to derive
 the radial stellar density variation associated to the Perseus arm in the Galactic anticenter.
 We want to see whether an overdensity of young stars associated to this arm can be identified. 
The Galactic anticenter direction was chosen because a second step of the project will be 
to obtain radial velocities for a subset of stars
to study the velocity perturbation due to the arm. For this purpose, the anticenter
direction is optimal since corrections related to Galactic rotation are negligible in the radial velocity component. 
Then, accurate distances to individual stars are critical for detecting
 the radial location of an overdensity. 
Str\"{o}mgren $uvbyH\beta$ photometry is the natural system for obtaining them.
It is also optimal to characterize the outer galaxy with Perseus and Cygnus arms \citep[see][]{2003A&A...397..133R} as candidates for massive 
spiral arms. Both the strong interstellar extinction in the direction to the Galactic center and the fact that Sagittarius may not be a major
 arm strongly suggest using the anticenter direction as the better option. 

The current paper presents the catalog. In Sect. \ref{survey} we describe the requirements for the survey, while in Sect. \ref{observations}
the data reduction procedures followed to obtain the photometry are explained, as well as the calibrations.
Section \ref{final} describes the steps in deriving the mean values in the final catalog from the individual measurements. 
Internal accuracies, limiting magnitudes, and 
comparisons with other catalogs can also be found in this section. Finally, Sect. \ref{summary} summarizes the characteristics of the survey and 
briefly discusses the next steps of the project.

\section{The Str\"{o}mgren anticenter survey}\label{survey}
The Str\"{o}mgren anticenter survey must fulfill several requirements in order to
 have the capability of detecting a possible overdensity of young stars induced by the Perseus arm,  
expected to be at about 2\,kpc \citep{2006Sci...311...54X} in the anticenter direction. 
Important requirements are  
1) a limiting magnitude to allow the detection of young stars up to about 3 kpc,
2) a survey area large enough to include a statistically significant number of young stellar objects 
in the anticenter direction, and
3) precise photometry to derive Str\"{o}mgren photometric distances beyond the Perseus arm. 

Stars with ages between 150 and 500 Myr (such as those with spectral types B5-A3) are the best population to study the possible
overdensity due to the Perseus spiral arm, since
they are young enough to still have a small intrinsic velocity dispersion 
(making them respond stronger to a perturbation),
but they are also old enough to have approached a dynamic equilibrium with the spiral perturbation.
Str\"{o}mgren $uvbyH\beta$ photometry \citep{1966ARA&A...4..433S}
is the natural system to identify this population, and it allows us to obtain accurate estimates of
individual distances and ages.

A statistically significant amount of stars in the photometric survey is needed, and they need to
reach at least 3\,kpc
from the Sun, so the required limiting magnitude is V=16$\fm$6 to detect an A0V star and V=17$\fm$7 for an A3V
star (assuming Av=1\,mag/kpc).
To select the survey area needed, Besan\c{c}on galaxy model simulations were used
\citep{2003A&A...409..523R}. Since it is likely that our galaxy has a relative
weak perturbation (i.e. $\sim$10\% variation in the disk density), approximately 900 B5-A3
 stars per radial 1\,kpc bin are needed for a 3$\sigma$ detection. 
Following the simulations, an 8$\degr ^2$ area is needed  to achieve
this number of stars. This is what we call the central part of the survey.
Since the volume covered in the nearby bins is small and also since there are saturation
effects, the statistics for these bins are too small, so an
extra area with brighter limiting magnitude was added, increasing the
survey  to 16$\degr ^2$. This surrounding area is named the outer part of the survey.

Distances that are more accurate than 25\% are needed to 
identify a 500\,pc spiral arm perturbation at 2\,kpc distance. This requirement 
imposes an upper limit on the errors in the absolute magnitude and in the interstellar extinction, both 
parameters to be 
derived from the Str\"{o}mgren photometry. The procedures
proposed by
 \cite{1978AJ.....83...48C}, \cite{1966ARA&A...4..433S}, and \cite{1979AJ.....84.1858C} allow us to compute
the absolute magnitudes for early (B0-A0), intermediate (A0-A3), and late 
type (A3-F0) stars, respectively. 
We estimated by simple error propagation that errors in $H\beta$ and $c_0$ smaller than 
0$\fm$020 and  0$\fm$035 result in distance errors between 25-15\% for B5-A0 stars 
and 25\% for A3 type stars. These values were computed assuming
an error smaller than 0$\fm$2 for the visual extinction ($A_V$).
  Although playing a 
role in the classification process, the error in other indexes makes no significant 
contribution to the estimation of distance errors.

Due to the Galactic warp, the Galactic plane is expected to be slightly
below Galactic latitude b=0$\degr$ in the anticenter direction \citep[see][]{2006A&A...451..515M}.
For that reason the center of the survey area was shifted down to
 b$\sim$-0$\fdg$5 in a low-extinction region \citep{2007MNRAS.378.1447F}.

\section{Observations and data reduction}\label{observations}
\subsection{Layout of the observations}
The observations were conducted using the Wide Field Camera (WFC) at the Isaac
Newton Telescope (INT) located at El Roque de los Muchachos in the Canary
Islands. The WFC is a four-chip mosaic of thinned AR-coated EEV 4K$\times$2K devices
with pixels size of 0$\farcs$333 and an edge to edge limit of the mosaic of 34$\farcm$2. The six filters used were Str\"{o}mgren $u$, $v$,
$b$, $y$, $H\beta _w$, $H\beta_n$ (see the central wavelength and band width in
Table \ref{filters}). Pixel binning of 1$\times$1 and slow read-out mode were used for the
observations, with a typical seeing of 1$\arcsec$-1$\farcs$5.
\begin{table}
\caption{Central wavelength and FWHM of the filters used.}\label{filters}
\centering
\begin{tabular} {c|ccccccc}
 & u & v & b & y & H$\beta _w$ & H$\beta _n$ \\ \hline
Central $\lambda$ (nm) & 348.0 & 411.0 & 469.5 & 550.5 & 486.1 & 486.1 \\ 
FWHM  (nm) & 33 & 15 & 21 & 24 & 17 & 3 \\ 
\end{tabular}
\end{table}
The WFC is the only
wide-field facility in the northern hemisphere that offers the full set of
Str\"{o}mgren filters.

\begin{figure}
\resizebox{\hsize}{!}{\includegraphics{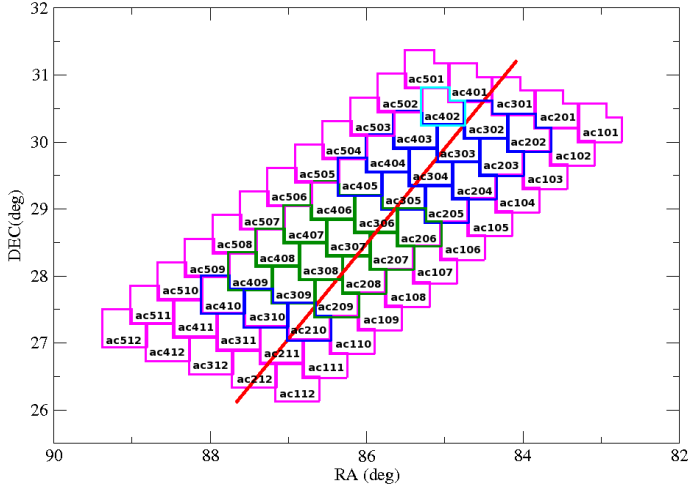}}
 \caption{Plot of the 60 WFC fields observed. Red line shows the b=-0$\fdg$5
plane. In green:
central fields observed during 2009A run. Dark blue: central fields observed during 2010B run.
Light blue: a central field observed during the 2011A run with a longer exposure time. Pink: outer
fields
observed during 2011A run with shorter exposure times and a single observation
per field. The anticenter fields are named 
ac$ij$, where $i$=1,...,5, and $j$=01,...,12.}\label{fields}
\end{figure}

Data from three different observing runs (2009A, 2010B, and 2011A) were used
for the catalog, and data from 2010A were excluded due to cloudy conditions.
 We were also granted some
director's discretionary time (in 2009 and 2011), but owing to bad weather these
nights
were not successful. Our 16$\degr ^2$ observing area was divided in a grid of 5$\times$12 WFC fields 
(see Fig. \ref{fields}), with an overlap
between them of 3$\arcmin$ in order to check for field-to-field variations. 

A different observational strategy was followed for the central and outer regions. 
For each of the 27 central fields (see Fig. \ref{fields}), three consecutive observations
 were obtained, 
with a shift of 10$\arcsec$ between them, in order to detect cosmic rays and avoid bad pixels.
 Exposure times for each filter and observation are detailed in Table \ref{obs}.
 The observations in the outer region, which includes 33 WFC fields, were planned to increase the 
statistics for nearby stars in the first kiloparsec distance
  bins. A single observation with shorter exposure times was conducted for 
 each of these fields (instead of three observations with offsets as in the central region).

\begin{table*}
\caption{Dates for the observing runs and exposure times for each observation and filter for our program fields.  
}\label{obs}
\centering
\begin{tabular} {c|c|c|c|cccccc|c}
Run & Dates & Photometric & AC fields & \multicolumn{6}{|c|}{Exposure times (s)} &
Calibration \\
  &   & nights & observed & $u$ & $v$ & $b$ & $y$ & $H\beta_w$ & $H\beta_n$ & fields \\\hline
2009A & 2009 Feb 12-16 & 2 & 12 & 100 & 40 & 40 & 30 & 40 & 200 & NGC1893,
ac308\\\hline
2010B & 2011 Jan 08-11 & 4 & 14 & 300 & 80 & 40 & 40 & 40 & 200 & NGC1893,
ac308, ac406, Praesepe \\\hline
2011A & 2011 Feb 16-17 & 2 & 1 & 720 & 120 & 100 & 100 & 100 & 720 & ac406,
Coma Berenices \\
  &   &   & 33  & 120 & 40 & 30 & 30 & 30 & 120 &  \\
\end{tabular}
\end{table*}

The calibration fields used for the transformation to the standard system are given in Table \ref{obs}. 
The open cluster NGC1893 was observed several times during the first two runs. 
The central part of this cluster has Str\"{o}mgren photometry available from \cite{1991MNRAS..253..649T} and \cite{2001AJ....121.2075M}.
To better control the transformation for each of the four WFC chips, several observations of this cluster were done, each time placing the 
center of the cluster in the center of each of the WFC chips. This strategy ensured that enough bright stars were available in each field, 
with about 50 stars per chip for the transformation to the standard system. These data were also used to calibrate other 
stars around the cluster, which were used as secondary standards. 
During the first observing nights, two anticenter fields were observed repeatedly (namely ac308 and ac406). 
After their calibration using NGC1893, they were used as deeper secondary standard fields in the following observing nights. 
The Coma Berenices \citep{1993RMxAA..25..129P,1969AJ.....74..407C} and Praesepe \citep{1969AJ.....74..818C,1991A&AS...90...25R}
open clusters were used as standard fields in some of the runs 
because they are older than NGC1893.

\subsection{Image pre-reduction and photometry extraction}
The images were reduced using several \textit{IRAF}\footnote{IRAF is
distributed by the National Optical Astronomy Observatories,
    which are operated by the Association of Universities for Research
    in Astronomy, Inc., under cooperative agreement with the National
    Science Foundation \citep{1986SPIE..627..733T}.} tasks. First, original
files were split into four different images, one for each chip, and the bias
derived from the 
overscan areas was subtracted. Bad pixels were replaced  by linear 
interpolation 
using the nearest good pixels through the \textit{fixpix} task. 
The linearity correction proposed on the CASU INT web 
page\footnote{http://www.ast.cam.ac.uk/$\sim$wfcsur/technical/foibles/index.php} was
applied, as was
the transformation factor from ADUs to electrons given in the manual.
 Flatfielding was applied using the sky flats obtained during
the observational runs. A mask was also applied to avoid the vignetted
corner of chip 3.
 
All the stars available in
the images were located using the \textit{daofind} routine.
Using the PSF photometry to derive instrumental magnitudes was carefully investigated. But the 
high dependency on the parameters that define the quality of a particular photometric image 
(seeing, sky background, etc.) can lead to differences on the order of 0$\fm$02 to 0$\fm$03, 
so the PSF fitting method was rejected and the full survey was reduced using a homogeneous 
aperture-corrected photometry. Twelve different aperture radii provided
twelve different magnitudes for each star. The \textit{daogrow} algorithm was used to obtain the aperture corrections and
the fitted radii. The final instrumental magnitudes were computed from the integration of the derived curve of growth.

\subsection{Extinction correction}
Our calibration fields (see Table \ref{obs}) were observed at several airmasses each night. Fitting their
differences in magnitude vs. the differences in airmass, the
extinction coefficients for each night were obtained. 
An intermediate range of
magnitudes for these stars was selected for the fit, avoiding the brightest
and the faintest 
ones. The measurements with airmass differences smaller than 0.1 were also rejected. 
The
fitted function was: $x''_i-x''_j=k (\chi_i-\chi_j)$
for all the available pairs of measurements  $i \neq j$ from the same star where $\chi$ is the airmass for each
measurement, and $x''$ indicates instrumental magnitudes.
 The extinction coefficients $k_x$ for each of the six filters ($u$, $v$, $b$, $y$,
H$\beta w$, and $H\beta n$) and night are listed in Table \ref{extcoef}, along with
the ranges in airmass used. 
The extinction-corrected magnitudes and indexes ($x'$) were then computed as 
\begin{eqnarray}
 && x'=x''-\chi \cdot k_x .
\end{eqnarray}
In the case of the $H\beta$ extinction coefficients, we computed and applied the average of the  
values obtained for H$\beta w$ and $H\beta n$. (As known, they are centered on the same wavelength.) 
This coefficient was applied independently to each filter, which allowed us to take 
the change in airmass between both exposures into account.

\subsection{Transformation to the standard system}
The photometry from our calibration fields (see photometric ranges in Table \ref{calfield}) 
 was used to obtain the transformation coefficients 
to the standard Str\"{o}mgren system. Several equations with different color terms were checked
 in order to select those that minimize the errors and correlations between coefficients, and also to 
avoid insignificant terms. The selected set of equations were
\begin{subeqnarray}\label{transf}
&& y'-V_{\mathrm{cat}} = A_1 + B_1\cdot(b-y)_{\mathrm{cat}} \label{transf1}, \\
&& (b-y)' = A_2 + C_2\cdot(b-y)_{\mathrm{cat}} \label{transf2},\\
&& c_1' = A_3 + B_3 \cdot(b-y)_{\mathrm{cat}} + C_3 \cdot c_{1\mathrm{cat}}  \label{transf3},\\
&& (v-b)' = A_4 + B_4\cdot (b-y)_{\mathrm{cat}} + C_4\cdot (v-b) + D_4\cdot c_{1\mathrm{cat}}, \label{transf4}\\
&& H\beta ' = A_5 + B_5 \cdot (b-y)_{\mathrm{cat}} + C_5 \cdot (H\beta_{\mathrm{cat}}
-2.8),  \label{transf5}
\end{subeqnarray}
where the prime indicates instrumental extinction-corrected variables and the
subscript $_{\mathrm{cat}}$ indicates the
standard values. The fitting 
equation in $(v-b)$ was selected instead of $m_1$, because $m_1$ has a narrower
dynamical range for young stars
than $(v-b)$.
As discussed in Sect. \ref{final}, the magnitudes in some individual filters may be missing, 
especially in the $u$ filter due to the need for very long exposure times. 
In this case Eqs. \ref{transf4} and \ref{transf5} had to be modified to avoid
the $c_1$  or $(b-y)$ indexes like

\begin{subeqnarray}\label{transfbis}
&& (v-b)' = \tilde{A_4} + \tilde{B_4}
(b-y)_{\mathrm{cat}} + \tilde{C_4}\cdot(v-b), \label{bis4}\\
&& H\beta ' = \tilde{A_5} + \tilde{C_5}\cdot(H\beta
_{\mathrm{cat}} -2.8). \label{bis5}
\end{subeqnarray}
Equations \ref{bis4} and \ref{bis5} were only used when the initial Eqs.
\ref{transf4} and \ref{transf5}
could not be used because some exposure in an individual filter was missing.
The obtained coefficients from each night are listed in Tables \ref{transcoef}
and \ref{transcoefbis}. Chip 3 has a slightly different behavior than the others, as 
can be seen in coefficients $A3$ and $C5$, possibly because it is vignetted. The errors in the photometric indexes are computed from direct
error propagation of the coefficients and magnitudes. Correlations among extinction coefficients were taken into account, as well as 
correlations 
among coefficients for the transformation to the standard system. 
Since correlations between both sets have not been taken into account, our errors 
can be slightly overestimated. However, as our standards have a wide range in both airmasses and colors, the contribution for such correlations
should be small.

The positions in the J2000 coordinate system were determined using \textit{wcs} and a 
fifth-order polynomial taking USNO-A2 
\citep{1998AAS...1931..2003M} as reference catalog.

\subsection{Illumination correction}
\begin{figure}
 \resizebox{\hsize}{!}{\includegraphics{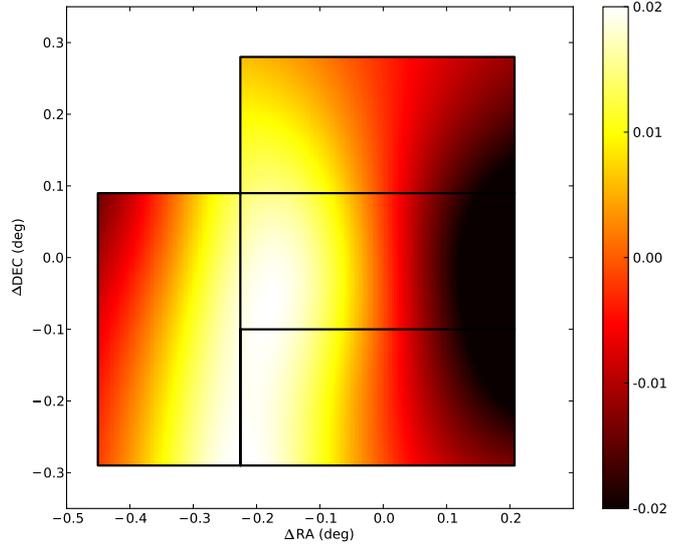}}
 \caption{Illumination differences in magnitudes at different field positions. Black lines show the location of the
four chips, and coordinates in degrees are centered on the central chip. The illumination differences at each position were computed as
the Gaussian-weighted mean (with $\sigma$=0.1$\degr$) of the residuals around each point.
}\label{IC}
\end{figure}

Since WFC covers a large field, the importance of the illumination
correction must be checked, owing to different illumination of the CCDs.
To do that, a field of stars was observed at several positions on the
CCDs, and only the $y$ filter was used since the illumination correction is not expected 
to be color dependent.
All the computed instrumental magnitudes for each star, corrected for atmospheric 
extinction, were averaged
to obtain a mean magnitude for each of them. 
The residuals between each individual magnitude and the computed mean magnitude were computed.
Figure \ref{IC} shows the smoothed distribution  for the residuals across the field of view of the WFC.
A weak trend in right ascension was found, reaching values up to +/-0$\fm$02, which is below our general photometric errors.
No significant trend was found in declination.
The scatter of the data did not justify using anything higher than a second-order fit in right ascension: $\Delta m=a\alpha^2+b\alpha+c$.
The results of the fit are $a$=-0.484$\pm$0.029\,mag/deg$^2$, $b$=-0.159$\pm$0.008\,mag/deg, $c$= 0.005$\pm$0.001\,mag  with residuals 
of 0.04\,mag.
This correction was applied to a test area with no significant change in the final mean magnitudes, 
except for a slight increase in the corresponding errors. Finally, the computed illumination correction was not applied.

\section{Final catalog}\label{final}
A catalog of 323794 individual measurements was compiled and
is available through the CDS (the detailed content for each column is described in Table \ref{readmemeas}).  
The astrometry for each individual measurement was computed as the mean of the coordinates derived from each of the filter images 
(six if all the magnitudes are available).
Next, a crossmatching process was executed, assuming that two or more measurements belong to the same target if their angular 
separation was smaller
than 3$\arcsec$. This crossmatch radius was selected as the value that minimizes  the number of outliers and maximizes 
the number of assignations, taking into account that it is around two to three times the size of the seeing.
\textit{STILTS}\footnote{Starlink Tables Infrastructure Library Tool Set, http://www.star.bris.ac.uk/$\sim$mbt/stilts/}
tools were used for that purpose, which also allowed us to assign an identifier (ID) to each star.
Finally a weighted mean was computed that yielded the final photometric indexes for each target.
Details of these computations are
1) those photometric indexes derived from magnitudes having FWHM smaller than two pixels
 were rejected (assumed to be bad pixels or wrong measurements);
2) a weighted mean was computed, where the weight applied was $w_i=1/\sigma_i^2$, and $\sigma_i$ is the 
individual error for each index, computed with full propagation errors;
3) outliers were rejected using a 5$\sigma$ rejection process, obtaining a final number of measurements different for each index 
(see Table \ref{Nmeast}); and
4) the weighted standard deviation and the error of the mean were computed for each index. 
The $m_1$ index was computed from the weighted mean of the individual $m_1$ measurements, so it
is not a direct linear combination of the mean $(v-b)$ and $(b-y)$ indexes. 
The final right ascension and declination coordinates for each target were also computed 
following a similar procedure.

Table \ref{Nmeast} shows the number of stars with 1, 2, 3, or more measurements. 
In the outer region, most stars have only one measurement, while in the central region, stars have usually three measurements, but
 six or nine if they were in an
overlap region. Stars in fields ac308 and ac406 were observed up to 20 times.
For the stars with a single measurement, the internal standard deviation computed by error propagation in Eqs. \ref{transf} and \ref{transfbis}
was assigned. For targets with two or more measurements, a flag indicating the coherence between them was computed for each index.
This flag gives the number of inconsistent pairs according to a Student's t-test (t $>$ 90\% was adopted). 
For the V magnitudes, 97\% of the stars with 
more than one measurement have a flag equal to zero; that is, all the measurements are consistent. Similar percentages are obtained for the 
other indexes.
 
The catalog with mean magnitudes and color indexes in the anticenter direction contains 96980 stars (also available through CDS), 
but not all of them 
have the full set of indexes. Table \ref{meas} shows the statistics of the final photometric data available.
A flag with six binary digits indicates the indexes available for each target. 
\begin{table}
 \caption{Number of stars for which mean magnitudes and indexes were computed using N individual measurements.}
\label{Nmeast}
\begin{tabular}{c|cccccc}
N  & $V$ & $(b-y)$ & $c_1$ & $(v-b)$ & $m_1$ & $H\beta$ \\ \hline
1 & 38740 & 38740 & 14353 & 24294 & 24294 & 33143 \\ 
2 & 12688 & 12705 & 5647 & 8320 & 8322 & 11358 \\ 
3 & 25859 & 25864 & 9139 & 15430 & 15429 & 22485 \\ 
$>$3 & 17968 & 17946 & 6985 & 11336 & 11335 & 15961\\ 
\end{tabular}
\end{table}
\begin{table}
\caption{Statistics of the number of stars as a function of the photometric information available.}
\label{meas}
\centering
\scriptsize
\begin{tabular}[!h]{c|cccccc|c}
$\sharp$ stars & $V$ & $(b-y)$ & $c_1$ & $(v-b)$ & $m_1$ & $H\beta$ & flagIA \\ \hline
1725 & - & - & - & - & - & $\times$ & 000001 \\ 
13259 & $\times$ & $\times$ & - & - & - & - & 110000 \\ 
624 & $\times$ & $\times$ & - & $\times$ & $\times$ & - & 110110 \\ 
22632 & $\times$ & $\times$ & - & $\times$ & $\times$ & $\times$ & 110111 \\ 
22616 & $\times$ & $\times$ & - & - & - & $\times$ & 110001 \\ 
150 & $\times$ & $\times$ & $\times$ & $\times$ & $\times$ & - & 111110 \\ 
35974 & $\times$ & $\times$ & $\times$ & $\times$ & $\times$ & $\times$ & 111111 \\ \hline
96980 & 95255 & 95255 & 36124 & 59380 & 59380 & 82947 & \\ 
\end{tabular}
\end{table}
As can be seen in Table \ref{meas} the catalog contains 35974 stars with all available indexes. 
Nonetheless, it is important to emphasize that it contains, in addition, 22632 stars with all indexes except $c_1$, 
22616 stars with  $V$, $(b-y)$ and $H\beta$, etc.
Reading the last row in Table \ref{meas}, it can be seen that we have about $\sim$6x10$^4$ stars with $m_1$ measurements or $\sim$8x10$^4$
stars with $H\beta$ index. 

Table \ref{reg} shows the spectral type distribution obtained by applying the procedure described in \cite{1991A&AS...87..319F} 
to the set of 35974 stars with all photometric indexes. This distribution can be compared with the contents of the \cite{1998A&AS..129..431H}
catalog, a local volume sample. As expected, our catalog contains a higher percentage of targets belonging to the late type group 
due to the different limiting magnitude. No stars in common were found between the two catalogs. 
Our survey area overlaps with the area covered by the North Hemisphere 
IPHAS survey \citep{2008MNRAS.388...89G}, with 
54109 stars in common. This overlap between both catalogs will be helpful for detecting stars with emission lines and peculiar features.

\begin{table}
\caption{Fraction of stars in the catalog for each spectral type}\label{reg}
\centering
\begin{tabular} {c|c|c|c|c}
 Early type & Intermediate type & \multicolumn{3}{|c}{Late type}\\
  B0-A0 & A0-A3 & A3-F0 & F0-G0 & G0 $\rightarrow$  \\ \hline
12\% & 8\% & 18\% & 56\% & 6\%  \\ 
\end{tabular}
\end{table}

GSC2 ID is also provided in the catalog and only $\sim$2\% of 
the stars do not have GSC counterparts.
In Table \ref{cat}, the first ten lines of the catalog are
provided, with the description of all the columns in Table \ref{readme}.

\subsection{Photometric precision}
\begin{figure}
 \resizebox{\hsize}{!}{\includegraphics{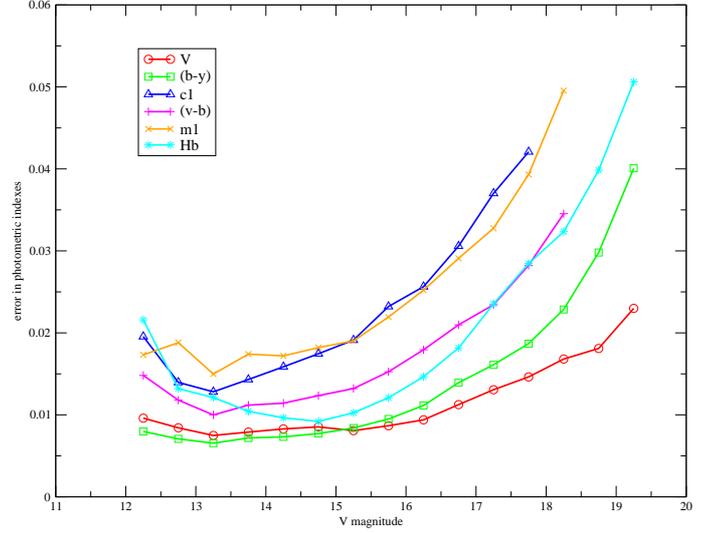}}
 \caption{Photometric precision, computed as the error of the mean, as a function of V magnitude for the stars with more than one measurement. Lines for the V magnitude and
the five standard indexes are plotted. Bins of 0$\fm$5 are used to compute the mean, and
inside each bin, outliers are rejected using a 5$\sigma$ clipping.
}\label{errindexNgt1}
\end{figure}

\begin{figure}
 \resizebox{\hsize}{!}{\includegraphics{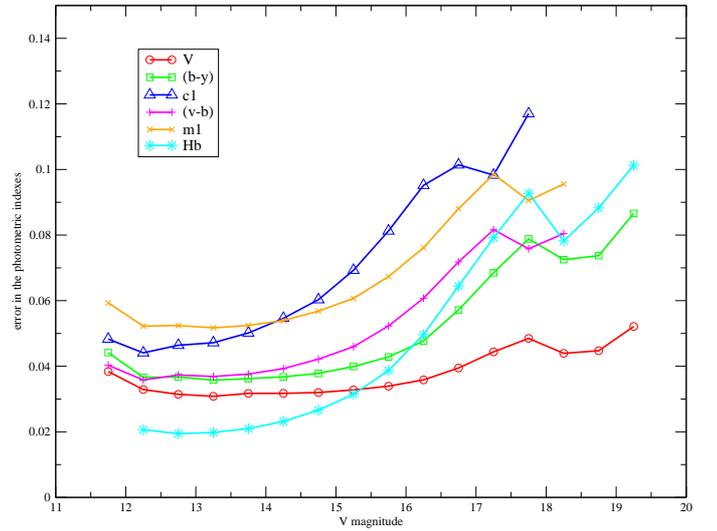}}
 \caption{Internal photometric standard deviation computed by error propagation in Eqs. \ref{transf} and \ref{transfbis},
as a function of V magnitude for those stars with only one measurement.
 Lines for the V magnitude and the five standard 
indexes are plotted. Bins of 0$\fm$5 are used in order to compute the mean, and
inside each bin, outliers are rejected using a 5$\sigma$ clipping.
}\label{errindexNeq1}
\end{figure}

The photometric precision obtained from the error of the mean in each of the indexes is shown in Fig.
\ref{errindexNgt1} as a function of V magnitude. 
For bright stars (V$<$16$^m$),
internal precisions below 0$\fm$01 were obtained for $V$ and $(b-y)$ and below 0$\fm$02
for the other indexes.
For fainter stars, the internal precision can reach up to 0$\fm$04-0$\fm$05.
For stars with a single measurement, the error of the mean could not be obtained, so we plot the internal standard deviation computed by 
error propagation in Eqs. \ref{transf} and \ref{transfbis}
(see Fig. \ref{errindexNeq1}). For stars 
brighter than $V$=12$^m$ the errors increase owing to saturation problems. 
The bump
 around V$\sim$16-17$^m$ is due to some nights with bright sky conditions, leading to larger error in the instrumental magnitudes.

The chip-to-chip variation was also checked using the stars observed several
times on different chips, 
i.e. in the overlap regions. Small variations were seen, but always less than the internal uncertainty.
The typical offsets between chips are smaller than 0$\fm$02.

\subsection{Astrometric precision}
\begin{figure}
 \resizebox{\hsize}{!}{\includegraphics{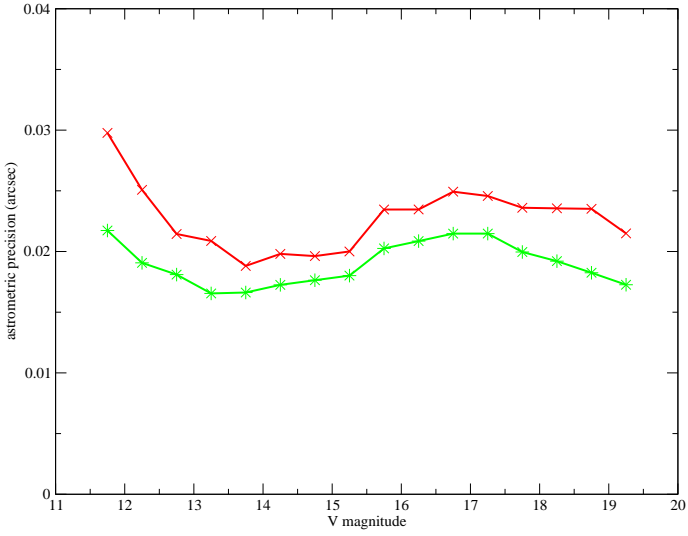}}
 \caption{Astrometric precision, computed as the error of the mean, as a function of V magnitude 
 in right ascension (in red and $\times$) and declination (in green and $\ast$)
computed for each
0$\fm$5 bin. Outliers (less than 5\%) were rejected using a 5$\sigma$ clipping in each bin.
}\label{astromaccur}
\end{figure}

\begin{figure}
 \resizebox{\hsize}{!}{\includegraphics{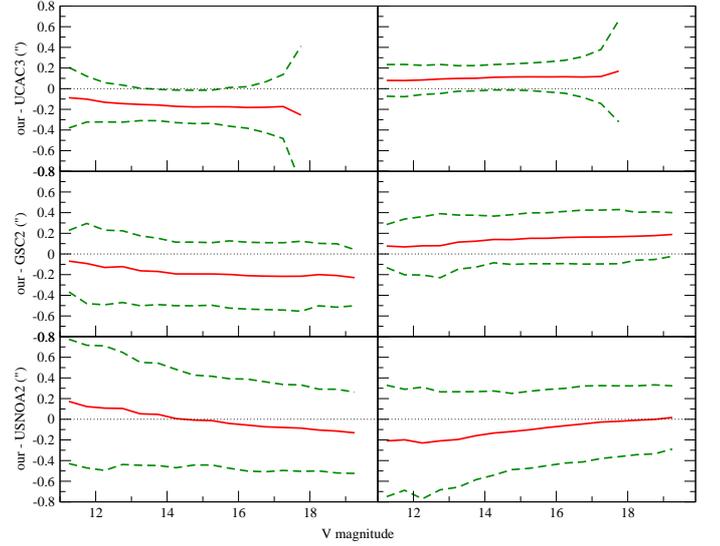}}
 \caption{Comparison of our astrometry with those from GSC2.3.2 (top), UCAC3
(middle), and 
USNO-A2 (bottom). Left: differences in $\alpha$cos$\delta$. Right: differences
in $\delta$. In red, mean differences. Green dashed lines show 1$\sigma$ ranges.
All differences are in arcsec.}\label{astrocomp}
\end{figure}

The internal astrometric precision, computed as the error of the mean, is around 0$\farcs$02 (see Fig. \ref{astromaccur}), 
less than one tenth of the pixel size (0$\farcs$333).
Figure \ref{astrocomp} shows the comparison between our astrometry with J2000.0 coordinates from UCAC3 \citep{2010AJ....139..2184Z},
GSC2.3.2 \citep{2008AJ....136..735L}, and USNO-A2.
Differences up to 0$\farcs$2 with UCAC3 and GSC2.3.2 can be observed, as well as a small trend in V magnitude, more pronounced in USNO-A2.
All these effects are explained by the different epochs of the three catalogs (1995-2000, 1988, and 1955 for UCAC3, GSC2, and USNO-A2, respectively).
USNO-A2 J2000.0 coordinates were used for the astrometric calibration 
because it contains stars fainter than UCAC3.
However, the mean epoch for USNO-A2 is 1955.0, and since proper motions are not available and cannot be taken into account, 
our coordinates do not contain the effect induced by the relative
Galactic rotation in the anticenter direction with respect to the Sun. This effect does not depend on the distance to the star 
(assuming a flat rotation curve)
and can reach 0$\farcs$2-0$\farcs$3 for differences in epoch of 50 years.
Figure \ref{astrocomp} shows that the dispersion increase from top to bottom,
again due to the differences between the epochs of our observations and catalog positions.
Furthermore, the decrease in the dispersion with increasing magnitude is explained by the effect of the intrinsic motion of the stars, 
stronger at short distances (so bright magnitudes). We used UCAC3 proper motions to check that the systematic trends disappear when
the difference in epochs (2010-1995) is considered. As mentioned, USNO-A2 does not provide proper motions, so the effects were not corrected in our final 
astrometric data. We verified that these trends have no effect on the crossmatching between our catalog and GSC2.3.2,
so the GSC ID is provided as additional information for the user.

 
\subsection{Limiting magnitude}
\begin{figure}
 \resizebox{\hsize}{!}{\includegraphics{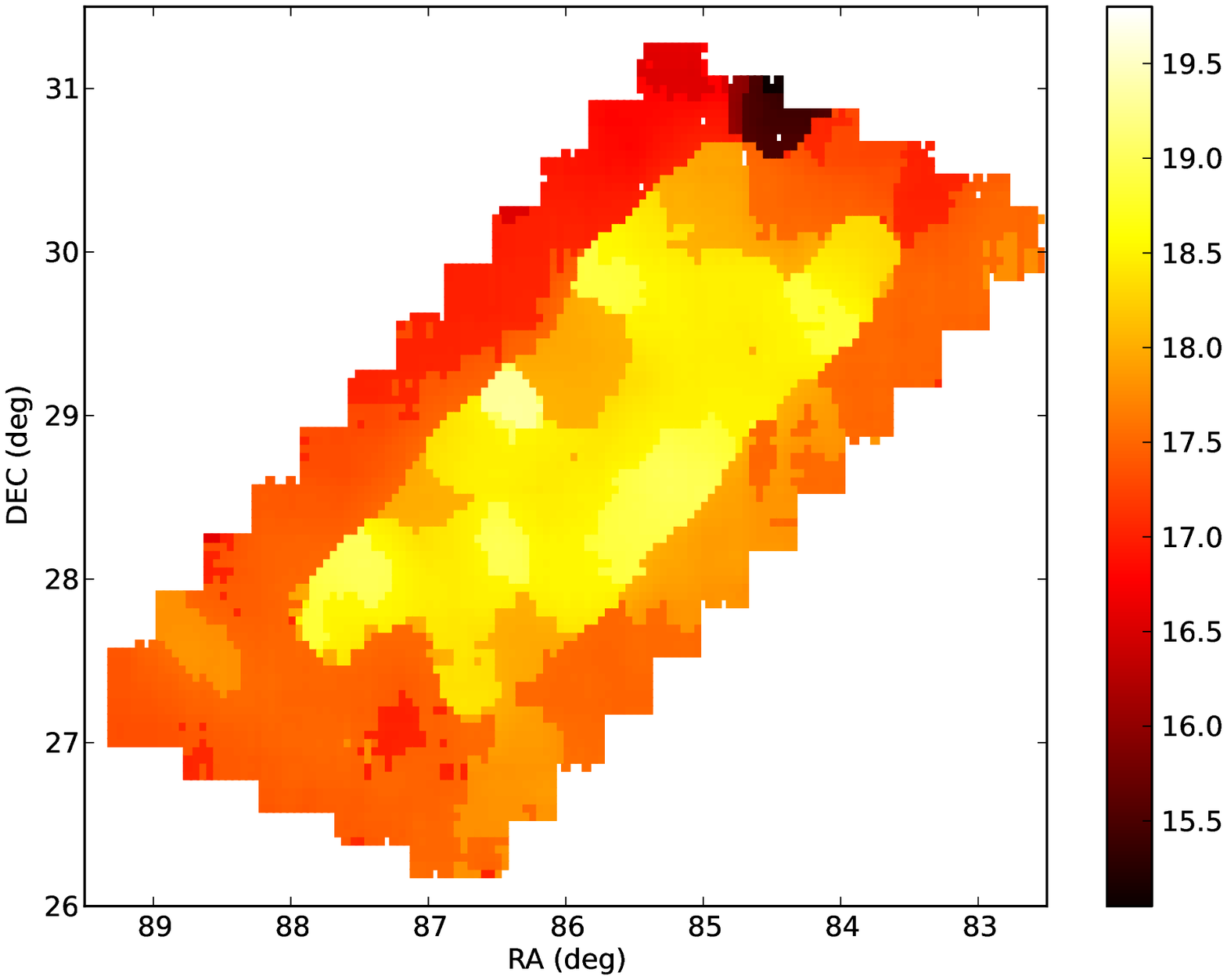}}
 \resizebox{\hsize}{!}{\includegraphics{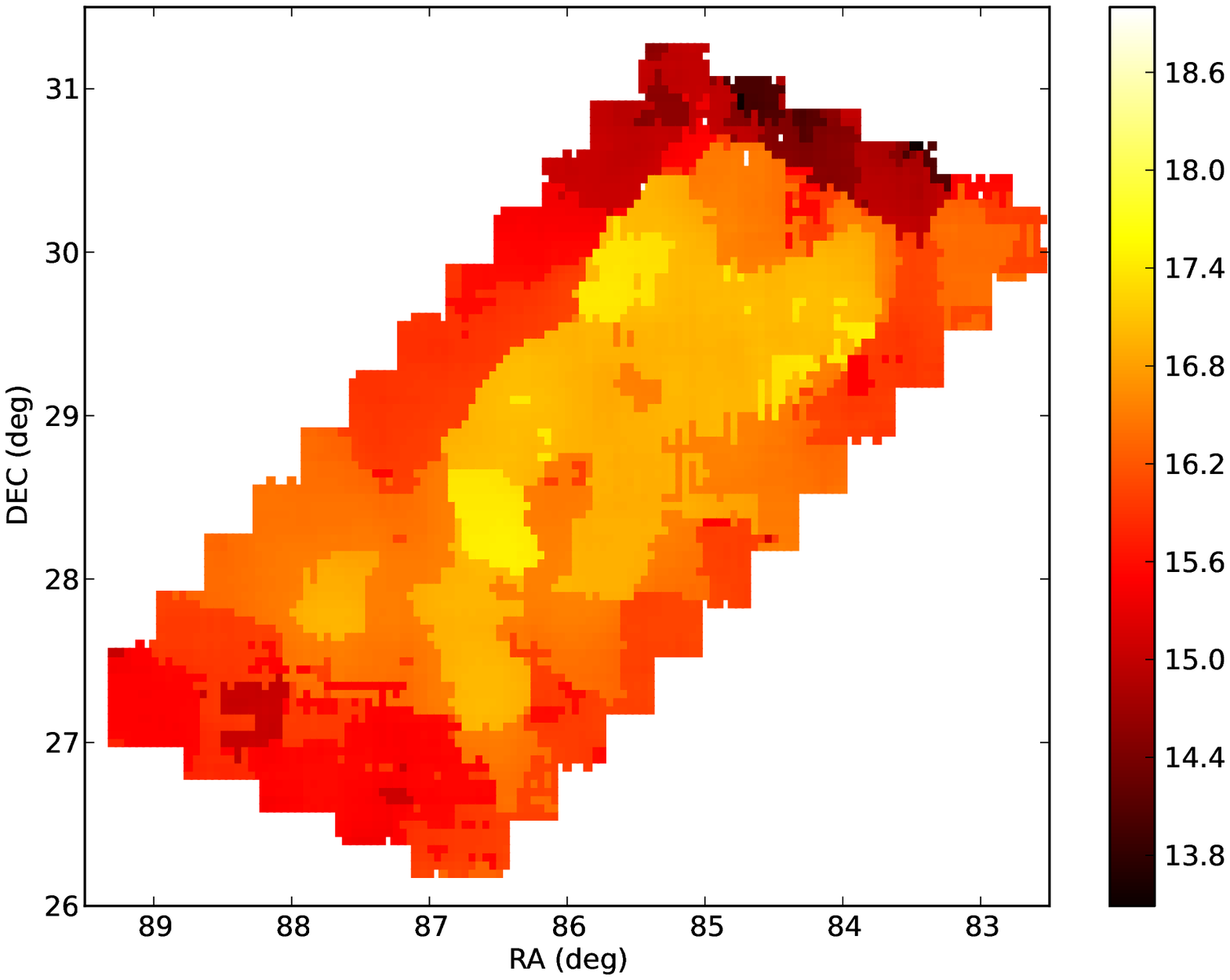}}
\caption{Two-dimensional distribution of the limiting magnitude showing the 90\%
completeness level. 
Top: Catalog of the 95255 stars with V magnitudes available. 
Bottom: Subcatalog of the 35974 stars with all the indexes available (See Table \ref{meas})}\label{limmag2D}
\end{figure}

The limiting magnitude was computed as the mean of the magnitudes at the peak star
counts in a magnitude histogram  and its two adjacent bins, before and after the peak, weighted by the number of stars in each bin.
We estimated that the limiting V magnitude computed with this simple algorithm provides the $\sim$90\% 
completeness limit. This was confirmed through the comparison of the V magnitude 
distribution of our catalog of stars with all available indexes with that of the full catalog containing
 all the stars with observed V magnitude (see Table \ref{meas}). 
This second catalog can be considered complete at the limiting magnitude of the previous one. 
The limiting magnitude obtained is not the same for all the survey. Data for  the outer area were obtained
using shorter exposure times. Also for the fields in the central region, the limiting 
magnitude is slightly variable due to both observation strategy and weather conditions. 
Figure \ref{limmag2D} shows its two-dimensional distribution. 
As mentioned, the catalog with all the available indexes is limited by the $u$ magnitude.
As can be seen in Fig. \ref{limmag2D}, our catalog of the 35974 stars with all indexes available reaches $\sim$90\% completeness at 
V$\sim$17$^m$ and V$\sim$15$\fm$5 for the central and outer regions, respectively. 

\begin{figure}
 \resizebox{\hsize}{!}{\includegraphics{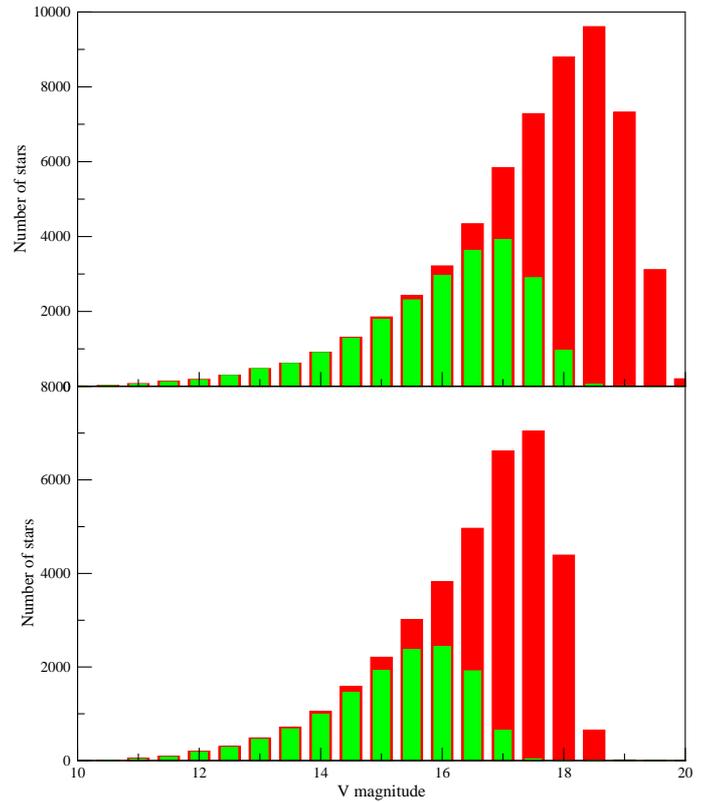}}
 \caption{V magnitude histogram. In red, stars with available V magnitudes. In
green, stars with all
indexes available. Top: Stars from the central deeper area. Bottom: Stars
from the outer area.}\label{limmag}
\end{figure}

Figure \ref{limmag} shows the V-magnitude histogram for the two main areas in our survey 
(the outer area and the central deeper region). 
In both cases the comparison between all the stars with available V magnitude and those with all
 indexes are provided.

\section{Summary and future work}\label{summary}
A catalog of Str\"{o}mgren photometry in the anticenter direction was built, 
covering a total area of 16$\degr ^2$, thereby providing photometric measurements for 96980 stars. 
The central 8$\degr ^2$ reach $\sim$90\% 
completeness at V$\sim$17$^m$, while the outer region
of $\sim$8$\degr ^2$, mostly observed with only one pointing, reaches this completeness
at $V\sim$15$\fm$5.
Photometric internal precisions between 0$\fm$01-0$\fm$02 for stars brighter than V=16$^m$ were obtained, increasing to 0$\fm$05 for some 
indexes and fainter stars (V$\sim$18-19$^m$).
The catalogs with the individual measurements and the final mean magnitudes and color indexes are published in electronic form via the CDS.

In a forthcoming paper the catalog will be used to determine whether there is an stellar overdensity of 
young stars induced by the Perseus spiral arm in the anticenter direction. 
That will allow us to fix the locus of the arm and to link this structure with the tracers observed in the Second and Third Galactic Quadrant.
The catalog is also being used to select a set of targets for a spectroscopic follow up. 
Radial velocities are being obtained using WYFFOS multiobject spectrograph installed at the WHT at the Canary Islands.
This will allow us to study the possible 
velocity perturbation due to the Perseus arm.
Undoubtedly this catalog will have other important scientific applications. 
As an example, it provides photometric metallicities for the FGK stars, from where a statistical analysis of the radial metallicity 
gradient can be undertaken. 
Our catalog could also be used to calibrate larger photometric surveys, such as IPHAS,
or for future Gaia spectroscopic follow up in the northern hemisphere.

\begin{acknowledgements}
We would like to thank the anonymous referee for his/her comments and suggestions that clearly helped to 
improve the paper.
This work was supported by the MINECO (Spanish Ministry of Economy) 
- FEDER through grant AYA2009-14648-C02-01 and CONSOLIDER CSD2007-00050.
M.Mongui\'{o} was supported by a Predoctoral fellowship from the Spanish Ministry (BES-2008-002471 through ESP2006-13855-C02-01 project).
\end{acknowledgements}

\bibliographystyle{aa}
\bibliography{Monguio}

\clearpage
\appendix
\section{Tables}\label{AppTables}
This appendix 
 includes some tables with detailed information about the catalog and the
photometric transformation process.
Table \ref{extcoef} shows the extinction 
coefficients obtained for each night and filter, as well as the airmass ranges covered by the calibration fields.
In Table 
\ref{calfield} we detail the range of magnitudes and color indexes covered by our calibration fields.
The coefficients for the transformation to the standard system through 
Eqs. \ref{transf} and \ref{transfbis} with their associated errors, for each of the observing nights,
are shown in Tables \ref{transcoef} and \ref{transcoefbis}, respectively.
Tables \ref{readmemeas} and \ref{readme} provide the column description for the two catalogs, which are
 individual measurements and final mean values, respectively.
Table \ref{cat} shows, as an example, the ten first rows of the mean catalog. For the individual measurements catalog, the ten first rows 
are not provided owing to the large amount of data included, but the full table can be found at the CDS.

\begin{table}[h]
\onecolumn \caption{Extinction coefficients obtained with the WFC/INT at El Roque de los Muchachos.
}\label{extcoef}
\centering
\begin{tabular} {c|cccccccc}
  & 2009 Feb 13 & 2009 Feb 17 & 2011 Feb 08 & 2011 Jan 09 & 2011 Jan 10 & 2011
Jan 11 & 2011 Feb 16 &
2011 Feb 17 \\\hline
$k_u$    & 0.553 & 0.575 & 0.605 & 0.574 & 0.556 & 0.537 & 0.550 & 0.528 \\
$k_v$    & 0.318 & 0.380 & 0.350 & 0.292 & 0.324 & 0.283 & 0.299 & 0.298 \\
$k_b$    & 0.219 & 0.245 & 0.261 & 0.195 & 0.214 & 0.174 & 0.206 & 0.232 \\
$k_y$    & 0.183 & 0.184 & 0.200 & 0.120 & 0.151 & 0.101 & 0.131 & 0.161 \\
$k_w$    & 0.191 & 0.213 & 0.218 & 0.148 & 0.174 & 0.129 & 0.168 & 0.163 \\
$k_n$    & 0.186 & 0.200 & 0.173 & 0.162 & 0.195 & 0.154 & 0.162 & 0.174 \\
$(k_w + k_n) /2$ & 0.188 & 0.205 & 0.195 & 0.155 & 0.184 & 0.141 & 0.165 & 0.169 \\
$\chi_{min} - \chi_{max}$&1.0 - 1.5&1.0 - 2.1&1.0 - 1.8&1.0 - 1.7&1.0 - 1.8&1.0 - 1.7&1.0 - 1.6&1.0 - 1.7\\
\end{tabular}
\tablefoot{Errors are between 0.01 and 0.03 magnitudes per airmass unit. 
Last row shows the airmass ranges used to derive these coefficients.}
\end{table}

\begin{table}[h]
\onecolumn \caption{Color index photometric ranges for the calibration fields}\label{calfield}
\centering
\begin{tabular} {c|cccccc}
Field& V & (b-y) & $c_1$ & (v-b) & $m_1$ & H$\beta$\\ \hline
Praesepe& 6 - 14 & 0.0 - 0.6 & 0.1 - 1.1 & 0.2 - 1.2& 0.1 - 0.6 & 2.5 - 2.9 \\ 
NGC1893& 13 - 16 & 0.1 - 1.2 & 0.05 - 1.25 & 0.3 - 1.6 & -0.15 - 0.7 & 2.4 - 3.0 \\
Coma& 5 - 11 & 0.1 - 1.1& 0.2 - 1.1 &0.2 - 1.6 & 0.1 - 0.6& 2.5 - 2.9 \\ 
ac308 & 13 - 16 & 0.25 - 1.6 & 0.03 - 1.3 &0.4 - 1.6& -0.2 - 0.7 & 2.5 - 3.0 \\
ac406& 13.5 - 16.5 & 0.2 - 1.4 & 0.1 - 1.3 &0.4 - 1.6& -0.1 - 0.7& 2.5 - 3.0 \\ 

\end{tabular}
\end{table}

\begin{landscape}
\begin{table}\scriptsize
\caption{Standard transformation coefficients}\label{transcoef}
\centering
\begin{tabular}
{c|cc|cc|ccc|cccc|ccc}
 & \multicolumn{2}{c|}{\textbf{Equation \ref{transf1}}} &
\multicolumn{2}{c|}{\textbf{Equation \ref{transf2}}} &
\multicolumn{3}{c|}{\textbf{Equation \ref{transf3}}} &
\multicolumn{4}{c|}{\textbf{Equation \ref{transf4}}} & 
\multicolumn{3}{c}{\textbf{Equation  \ref{transf5}}}
\\\hline
 chip& $A_1$ & $B_1$ & $A_2$ & $C_2$ & $A_3$ & $B_3$ & $C_3$ & $A_4$ & $B_4$ & $C_4$ & $D_4$ & $A_5$ & $B_5$ & $C_5$  \\\hline
\multicolumn{15}{c}{\textbf{2009 Feb 13}} \\\hline
 1 & -24.909$\pm$0.001 & -0.073$\pm$0.002 & -0.225$\pm$0.001 & 0.975$\pm$0.003 &
-0.323$\pm$0.006 & -0.072$\pm$0.008 & 0.950$\pm$0.004 & 0.169$\pm$0.003 &
0.028$\pm$0.003 & 0.969$\pm$0.007 & 0.152$\pm$0.009 & -2.330$\pm$0.002 &
0.034$\pm$0.003 & 0.843$\pm$0.004 \\
 2 & -24.704$\pm$0.001 & -0.056$\pm$0.002 & -0.266$\pm$0.001 & 0.967$\pm$0.002 &
-0.332$\pm$0.005 & -0.126$\pm$0.006 & 0.975$\pm$0.004 & 0.226$\pm$0.003 &
0.013$\pm$0.002 & 0.940$\pm$0.006 & 0.263$\pm$0.007 & -2.328$\pm$0.001 &
0.046$\pm$0.003 & 0.873$\pm$0.004 \\ 
 3 & -24.853$\pm$0.001 & -0.073$\pm$0.002 & -0.238$\pm$0.001 & 0.989$\pm$0.002 &
0.013$\pm$0.004 & -0.103$\pm$0.005 & 0.954$\pm$0.004 & 0.285$\pm$0.002 &
0.017$\pm$0.002 & 0.957$\pm$0.004 & 0.197$\pm$0.006 & -2.343$\pm$0.001 &
0.047$\pm$0.002 & 0.901$\pm$0.004 \\ 
 4 & -24.796$\pm$0.001 & -0.078$\pm$0.001 & -0.217$\pm$0.001 & 0.971$\pm$0.002 &
-0.177$\pm$0.004 & -0.079$\pm$0.006 & 0.982$\pm$0.003 & 0.343$\pm$0.002 &
0.015$\pm$0.002 & 0.892$\pm$0.005 & 0.298$\pm$0.006 & -2.332$\pm$0.001 &
0.045$\pm$0.002 & 0.857$\pm$0.003 \\ \hline 
\multicolumn{15}{c}{\textbf{2009 Feb 16}} \\\hline
 1 & -24.827$\pm$0.001 & -0.067$\pm$0.002 & -0.244$\pm$0.002 & 0.974$\pm$0.003 &
-0.241$\pm$0.005 & -0.153$\pm$0.006 & 0.971$\pm$0.005 & 0.133$\pm$0.005 &
0.001$\pm$0.005 & 0.918$\pm$0.009 & 0.220$\pm$0.010 & -2.332$\pm$0.001 &
0.047$\pm$0.002 & 0.856$\pm$0.004 \\ 
 2 & -24.626$\pm$0.002 & -0.050$\pm$0.003 & -0.289$\pm$0.002 & 0.982$\pm$0.003 &
-0.229$\pm$0.007 & -0.210$\pm$0.008 & 0.960$\pm$0.006 & 0.167$\pm$0.004 &
0.024$\pm$0.004 & 0.931$\pm$0.008 & 0.287$\pm$0.009 & -2.320$\pm$0.002 &
0.080$\pm$0.003 & 0.844$\pm$0.005 \\ 
 3 & -24.767$\pm$0.002 & -0.087$\pm$0.003 & -0.247$\pm$0.002 & 0.981$\pm$0.003 &
0.094$\pm$0.007 & -0.118$\pm$0.008 & 0.982$\pm$0.006 & 0.238$\pm$0.004 &
0.014$\pm$0.004 & 0.943$\pm$0.008 & 0.205$\pm$0.010 & -2.350$\pm$0.002 &
0.036$\pm$0.003 & 1.003$\pm$0.005 \\ 
 4 & -24.718$\pm$0.002 & -0.069$\pm$0.003 & -0.248$\pm$0.002 & 0.985$\pm$0.003 &
-0.075$\pm$0.008 & -0.132$\pm$0.010 & 0.961$\pm$0.006 & 0.292$\pm$0.004 &
0.016$\pm$0.003 & 0.946$\pm$0.007 & 0.266$\pm$0.009 & -2.322$\pm$0.002 &
0.044$\pm$0.004 & 0.843$\pm$0.005 \\ \hline
\multicolumn{15}{c}{\textbf{2011 Jan 08}} \\\hline
 1 & -24.807$\pm$0.001 & -0.068$\pm$0.002 & -0.216$\pm$0.002 & 0.968$\pm$0.002 &
-0.304$\pm$0.005 & -0.147$\pm$0.006 & 0.977$\pm$0.005 & 0.181$\pm$0.003 &
0.010$\pm$0.003 & 0.940$\pm$0.005 & 0.201$\pm$0.005 & -2.330$\pm$0.001 &
0.043$\pm$0.002 & 0.866$\pm$0.005 \\ 
 2 & -24.591$\pm$0.001 & -0.054$\pm$0.002 & -0.264$\pm$0.002 & 0.991$\pm$0.003 &
-0.311$\pm$0.005 & -0.144$\pm$0.006 & 0.966$\pm$0.005 & 0.219$\pm$0.003 &
0.033$\pm$0.003 & 0.957$\pm$0.005 & 0.236$\pm$0.006 & -2.324$\pm$0.002 &
0.064$\pm$0.003 & 0.875$\pm$0.005 \\ 
 3 & -24.696$\pm$0.001 & -0.087$\pm$0.002 & -0.210$\pm$0.002 & 0.983$\pm$0.003 &
-0.017$\pm$0.006 & -0.128$\pm$0.007 & 0.974$\pm$0.005 & 0.281$\pm$0.003 &
0.023$\pm$0.003 & 0.952$\pm$0.005 & 0.200$\pm$0.006 & -2.323$\pm$0.002 &
0.042$\pm$0.003 & 0.998$\pm$0.006 \\ 
 4 & -24.693$\pm$0.002 & -0.073$\pm$0.003 & -0.224$\pm$0.002 & 0.988$\pm$0.004 &
-0.161$\pm$0.008 & -0.115$\pm$0.010 & 0.948$\pm$0.006 & 0.335$\pm$0.003 &
0.021$\pm$0.002 & 0.965$\pm$0.005 & 0.246$\pm$0.007 & -2.306$\pm$0.002 &
0.039$\pm$0.004 & 0.847$\pm$0.005 \\\hline
\multicolumn{15}{c}{\textbf{2011 Jan 09}} \\\hline
 1 & -24.735$\pm$0.005 & -0.069$\pm$0.012 & -0.231$\pm$0.005 & 0.970$\pm$0.013 &
-0.322$\pm$0.012 & -0.139$\pm$0.028 & 0.985$\pm$0.013 & 0.165$\pm$0.005 &
0.011$\pm$0.005 & 0.951$\pm$0.021 & 0.192$\pm$0.025 & -2.324$\pm$0.007 &
0.048$\pm$0.017 & 0.860$\pm$0.025\\ 
 2 & -24.514$\pm$0.005 & -0.053$\pm$0.013 & -0.281$\pm$0.005 & 0.988$\pm$0.013 &
-0.328$\pm$0.015 & -0.167$\pm$0.034 & 0.971$\pm$0.015 & 0.210$\pm$0.006 &
0.029$\pm$0.006 & 0.955$\pm$0.024 & 0.240$\pm$0.029 & -2.322$\pm$0.007 &
0.067$\pm$0.018 & 0.882$\pm$0.026 \\
 3 & -24.621$\pm$0.010 & -0.087$\pm$0.025 & -0.232$\pm$0.008 & 0.980$\pm$0.019 &
-0.036$\pm$0.011 & -0.112$\pm$0.025 & 0.973$\pm$0.012 & 0.269$\pm$0.005 &
0.026$\pm$0.005 & 0.951$\pm$0.020 & 0.203$\pm$0.025 & -2.326$\pm$0.006 &
0.036$\pm$0.016 & 1.015$\pm$0.026 \\ 
 4 & -24.618$\pm$0.005 & -0.072$\pm$0.013 & -0.243$\pm$0.005 & 0.990$\pm$0.013 &
-0.190$\pm$0.013 & -0.122$\pm$0.030 & 0.956$\pm$0.014 & 0.328$\pm$0.006 &
0.019$\pm$0.006 & 0.972$\pm$0.023 & 0.234$\pm$0.029 & -2.303$\pm$0.007 &
0.037$\pm$0.018 & 0.854$\pm$0.027 \\ \hline
\multicolumn{15}{c}{\textbf{2011 Jan 10}} \\\hline
 1 & -24.749$\pm$0.007 & -0.109$\pm$0.019 & -0.216$\pm$0.009 & 0.956$\pm$0.022 &
-0.231$\pm$0.019 & -0.139$\pm$0.045 & 0.964$\pm$0.020 & 0.161$\pm$0.006 &
0.013$\pm$0.006 & 0.966$\pm$0.025 & 0.181$\pm$0.027 & -2.313$\pm$0.010 &
0.027$\pm$0.027 & 0.828$\pm$0.037 \\ 
 2 & -24.532$\pm$0.008 & -0.077$\pm$0.020 & -0.265$\pm$0.006 & 0.970$\pm$0.015 &
-0.254$\pm$0.019 & -0.149$\pm$0.043 & 0.974$\pm$0.020 & 0.190$\pm$0.009 &
0.012$\pm$0.010 & 1.036$\pm$0.033 & 0.170$\pm$0.042 & -2.327$\pm$0.008 &
0.009$\pm$0.020 & 0.953$\pm$0.032 \\ 
 3 & -24.646$\pm$0.008 & -0.074$\pm$0.020 & -0.237$\pm$0.009 & 0.985$\pm$0.024 &
0.055$\pm$0.014 & -0.141$\pm$0.033 & 0.960$\pm$0.015 & 0.273$\pm$0.008 &
0.013$\pm$0.008 & 0.975$\pm$0.026 & 0.166$\pm$0.036 & -2.323$\pm$0.011 &
0.057$\pm$0.028 & 0.905$\pm$0.041 \\ 
 4 & -24.638$\pm$0.009 & -0.087$\pm$0.022 & -0.229$\pm$0.007 & 0.971$\pm$0.016 &
-0.133$\pm$0.024 & -0.032$\pm$0.051 & 0.926$\pm$0.023 & 0.342$\pm$0.006 &
0.011$\pm$0.006 & 1.022$\pm$0.019 & 0.135$\pm$0.027 & -2.299$\pm$0.012 &
0.018$\pm$0.031 & 0.815$\pm$0.041 \\ \hline
\multicolumn{15}{c}{\textbf{2011 Jan 11}} \\\hline
 1 & -24.715$\pm$0.002 & -0.062$\pm$0.004 & -0.234$\pm$0.002 & 0.974$\pm$0.004 &
-0.260$\pm$0.006 & -0.148$\pm$0.008 & 0.973$\pm$0.007 & 0.165$\pm$0.004 &
0.002$\pm$0.004 & 0.945$\pm$0.007 & 0.188$\pm$0.008 & -2.340$\pm$0.003 &
0.053$\pm$0.005 & 0.837$\pm$0.009 \\ 
 2 & -24.494$\pm$0.003 & -0.047$\pm$0.004 & -0.282$\pm$0.003 & 0.988$\pm$0.004 &
-0.339$\pm$0.007 & -0.176$\pm$0.010 & 0.969$\pm$0.005 & 0.199$\pm$0.004 &
0.017$\pm$0.003 & 0.979$\pm$0.007 & 0.212$\pm$0.008 & -2.340$\pm$0.003 &
0.071$\pm$0.005 & 0.886$\pm$0.007 \\ 
 3 & -24.597$\pm$0.002 & -0.072$\pm$0.004 & -0.238$\pm$0.002 & 0.984$\pm$0.004 &
0.030$\pm$0.006 & -0.132$\pm$0.009 & 0.961$\pm$0.006 & 0.265$\pm$0.003 &
0.019$\pm$0.003 & 0.939$\pm$0.006 & 0.213$\pm$0.007 & -2.338$\pm$0.003 &
0.057$\pm$0.005 & 0.959$\pm$0.008 \\ 
 4 & -24.598$\pm$0.003 & -0.063$\pm$0.005 & -0.231$\pm$0.002 & 0.978$\pm$0.004 &
-0.138$\pm$0.007 & -0.087$\pm$0.009 & 0.933$\pm$0.005 & 0.327$\pm$0.003 &
0.015$\pm$0.003 & 0.971$\pm$0.007 & 0.213$\pm$0.008 & -2.314$\pm$0.003 &
0.045$\pm$0.005 & 0.828$\pm$0.007 \\ \hline
\multicolumn{15}{c}{\textbf{2011 Feb 16}} \\\hline
1 & -24.745$\pm$0.004 & -0.080$\pm$0.008 & -0.233$\pm$0.007 & 0.973$\pm$0.013 &
-0.325$\pm$0.016 & -0.087$\pm$0.026 & 0.971$\pm$0.015 & 0.189$\pm$0.009 &
0.014$\pm$0.009 & 0.981$\pm$0.017 & 0.129$\pm$0.023 & -2.307$\pm$0.007 &
0.024$\pm$0.013 & 0.831$\pm$0.018 \\ 
2 & -24.519$\pm$0.004 & -0.059$\pm$0.006 & -0.282$\pm$0.005 & 0.964$\pm$0.009 &
-0.299$\pm$0.018 & -0.144$\pm$0.025 & 0.933$\pm$0.011 & 0.228$\pm$0.010 &
0.027$\pm$0.007 & 0.972$\pm$0.020 & 0.203$\pm$0.023 & -2.314$\pm$0.006 &
0.046$\pm$0.011 & 0.883$\pm$0.012 \\ 
3 & -24.624$\pm$0.003 & -0.081$\pm$0.005 & -0.258$\pm$0.005 & 1.003$\pm$0.008 &
-0.023$\pm$0.014 & -0.100$\pm$0.021 & 0.960$\pm$0.011 & 0.291$\pm$0.007 &
0.021$\pm$0.006 & 0.957$\pm$0.017 & 0.175$\pm$0.020 & -2.323$\pm$0.005 &
0.053$\pm$0.009 & 0.950$\pm$0.013 \\ 
4 & -24.633$\pm$0.004 & -0.063$\pm$0.007 & -0.248$\pm$0.006 & 0.992$\pm$0.012 &
-0.210$\pm$0.019 & -0.052$\pm$0.030 & 0.955$\pm$0.013 & 0.354$\pm$0.010 &
0.018$\pm$0.007 & 1.010$\pm$0.023 & 0.152$\pm$0.026 & -2.293$\pm$0.007 &
0.040$\pm$0.014 & 0.823$\pm$0.013 \\ \hline
\multicolumn{15}{c}{\textbf{2011 Feb 17}} \\\hline
1 & -24.761$\pm$0.004 & -0.097$\pm$0.006 & -0.214$\pm$0.004 & 0.956$\pm$0.007 &
-0.293$\pm$0.010 & -0.157$\pm$0.014 & 0.983$\pm$0.011 & 0.206$\pm$0.006 &
0.018$\pm$0.007 & 0.966$\pm$0.011 & 0.173$\pm$0.014 & -2.294$\pm$0.004 &
0.049$\pm$0.007 & 0.849$\pm$0.011 \\ 
2 & -24.546$\pm$0.004 & -0.058$\pm$0.006 & -0.282$\pm$0.005 & 0.949$\pm$0.009 &
-0.307$\pm$0.015 & -0.174$\pm$0.020 & 0.954$\pm$0.009 & 0.259$\pm$0.009 &
0.021$\pm$0.006 & 0.912$\pm$0.015 & 0.300$\pm$0.018 & -2.303$\pm$0.004 &
0.054$\pm$0.008 & 0.871$\pm$0.008 \\ 
3 & -24.663$\pm$0.003 & -0.078$\pm$0.005 & -0.240$\pm$0.004 & 0.980$\pm$0.007 &
-0.005$\pm$0.013 & -0.147$\pm$0.017 & 0.940$\pm$0.010 & 0.320$\pm$0.006 &
0.027$\pm$0.005 & 0.973$\pm$0.011 & 0.183$\pm$0.014 & -2.311$\pm$0.004 &
0.056$\pm$0.007 & 0.936$\pm$0.009 \\ 
4 & -24.665$\pm$0.003 & -0.069$\pm$0.006 & -0.236$\pm$0.004 & 0.990$\pm$0.007 &
-0.171$\pm$0.014 & -0.112$\pm$0.021 & 0.948$\pm$0.011 & 0.386$\pm$0.009 &
0.015$\pm$0.007 & 1.005$\pm$0.018 & 0.173$\pm$0.018 & -2.275$\pm$0.005 &
0.027$\pm$0.008 & 0.846$\pm$0.010 \\ 
 \end{tabular}
\end{table}
\end{landscape}

\begin{table*}
\caption{Standard transformation coefficients for Eqs. \ref{transfbis}}\label{transcoefbis}
\centering
\begin{tabular}[h]{c|ccc|cc}\hline
  \textbf{chip} & \multicolumn{3}{c|}{\textbf{Equation \ref{bis4}}} &
\multicolumn{2}{c}{\textbf{Equation \ref{bis5}}}\\ \hline
& $\tilde{A_4}$ & $\tilde{B_4}$ & $\tilde{C_4}$ & $\tilde{A_5}$ & $\tilde{C_5}$ \\ \hline
 \multicolumn{6}{c}{\textbf{2009 Feb 13}} \\\hline
 1 & 0.195$\pm$0.002 & 0.940$\pm$0.007 & 0.165$\pm$0.009 & -2.315$\pm$0.001
& 0.865$\pm$0.004\\
 2 & 0.237$\pm$0.002 & 0.929$\pm$0.005 & 0.269$\pm$0.007 & -2.308$\pm$0.001
& 0.911$\pm$0.004\\
 3 & 0.300$\pm$0.001 & 0.956$\pm$0.004 & 0.185$\pm$0.006 & -2.325$\pm$0.001
& 0.957$\pm$0.004\\
 4 & 0.356$\pm$0.001 & 0.891$\pm$0.005 & 0.290$\pm$0.006 & -2.315$\pm$0.001
& 0.896$\pm$0.003\\ \hline
\multicolumn{6}{c}{\textbf{2009 Feb 16}} \\\hline
 1 & 0.135$\pm$0.003 & 0.917$\pm$0.008 & 0.222$\pm$0.010 & -2.306$\pm$0.001
& 0.882$\pm$0.005\\
 2 & 0.185$\pm$0.003 & 0.903$\pm$0.007 & 0.312$\pm$0.008 & -2.272$\pm$0.001
& 0.877$\pm$0.008\\
 3 & 0.252$\pm$0.003 & 0.930$\pm$0.008 & 0.213$\pm$0.009 & -2.332$\pm$0.001
& 1.033$\pm$0.005\\
 4 & 0.308$\pm$0.003 & 0.935$\pm$0.007 & 0.266$\pm$0.010 & -2.299$\pm$0.001
& 0.867$\pm$0.005\\ \hline
\multicolumn{6}{c}{\textbf{2011 Jan 08}} \\\hline
 1 & 0.189$\pm$0.002 & 0.931$\pm$0.004 & 0.208$\pm$0.005 & -2.292$\pm$0.002
& 0.886$\pm$0.011\\
 2 & 0.245$\pm$0.002 & 0.922$\pm$0.005 & 0.264$\pm$0.006 & -2.277$\pm$0.001
& 0.903$\pm$0.010\\
 3 & 0.303$\pm$0.002 & 0.929$\pm$0.005 & 0.212$\pm$0.006 & -2.295$\pm$0.001
& 1.029$\pm$0.009\\
 4 & 0.356$\pm$0.002 & 0.946$\pm$0.005 & 0.253$\pm$0.007 & -2.272$\pm$0.002
& 0.872$\pm$0.011\\ \hline
\multicolumn{6}{c}{\textbf{2011 Jan 09}} \\\hline
 1 & 0.178$\pm$0.002 & 0.969$\pm$0.005 & 0.174$\pm$0.007 & -2.302$\pm$0.001
& 0.891$\pm$0.004\\
 2 & 0.225$\pm$0.002 & 0.941$\pm$0.007 & 0.262$\pm$0.009 & -2.297$\pm$0.001
& 0.937$\pm$0.007\\
 3 & 0.292$\pm$0.002 & 0.960$\pm$0.006 & 0.185$\pm$0.008 & -2.302$\pm$0.001
& 0.979$\pm$0.005\\
 4 & 0.348$\pm$0.002 & 0.946$\pm$0.006 & 0.229$\pm$0.008 & -2.290$\pm$0.001
& 0.899$\pm$0.004\\ \hline
\multicolumn{6}{c}{\textbf{2011 Jan 10}} \\\hline
 1 & 0.172$\pm$0.002 & 0.948$\pm$0.005 & 0.189$\pm$0.006 & -2.308$\pm$0.001
& 0.882$\pm$0.004\\
 2 & 0.216$\pm$0.002 & 0.950$\pm$0.004 & 0.232$\pm$0.006 & -2.306$\pm$0.001
& 0.919$\pm$0.005\\
 3 & 0.283$\pm$0.002 & 0.946$\pm$0.005 & 0.196$\pm$0.006 & -2.306$\pm$0.001
& 0.978$\pm$0.004\\
 4 & 0.341$\pm$0.002 & 0.923$\pm$0.005 & 0.253$\pm$0.007 & -2.293$\pm$0.001
& 0.874$\pm$0.004\\ \hline
\multicolumn{6}{c}{\textbf{2011 Jan 11}} \\\hline
 1 & 0.166$\pm$0.002 & 0.943$\pm$0.006 & 0.189$\pm$0.007 & -2.312$\pm$0.002
& 0.854$\pm$0.010\\
 2 & 0.215$\pm$0.003 & 0.961$\pm$0.006 & 0.224$\pm$0.008 & -2.302$\pm$0.001
& 0.913$\pm$0.009\\
 3 & 0.281$\pm$0.002 & 0.921$\pm$0.005 & 0.225$\pm$0.007 & -2.308$\pm$0.001
& 0.983$\pm$0.009\\
 4 & 0.341$\pm$0.002 & 0.954$\pm$0.006 & 0.223$\pm$0.008 & -2.292$\pm$0.001
& 0.849$\pm$0.007\\ \hline
\multicolumn{6}{c}{\textbf{2011 Feb 16}} \\\hline
 1 & 0.198$\pm$0.007 & 0.969$\pm$0.016 & 0.138$\pm$0.022 & -2.295$\pm$0.003
& 0.841$\pm$0.017\\
 2 & 0.256$\pm$0.007 & 0.926$\pm$0.017 & 0.235$\pm$0.023 & -2.289$\pm$0.002
& 0.903$\pm$0.011\\
 3 & 0.308$\pm$0.006 & 0.931$\pm$0.015 & 0.197$\pm$0.020 & -2.296$\pm$0.002
& 0.967$\pm$0.013\\
 4 & 0.372$\pm$0.007 & 0.978$\pm$0.020 & 0.174$\pm$0.025 & -2.274$\pm$0.002
& 0.839$\pm$0.012\\ \hline
\multicolumn{6}{c}{\textbf{2011 Feb 17}} \\\hline
 1 & 0.216$\pm$0.005 & 0.955$\pm$0.010 & 0.182$\pm$0.013 & -2.282$\pm$0.002
& 0.861$\pm$0.012\\
 2 & 0.281$\pm$0.006 & 0.883$\pm$0.012 & 0.319$\pm$0.018 & -2.285$\pm$0.001
& 0.890$\pm$0.008\\
 3 & 0.343$\pm$0.005 & 0.946$\pm$0.010 & 0.203$\pm$0.014 & -2.293$\pm$0.001
& 0.957$\pm$0.010\\
 4 & 0.400$\pm$0.006 & 0.982$\pm$0.015 & 0.187$\pm$0.018 & -2.274$\pm$0.002
& 0.856$\pm$0.010\\
  \end{tabular}
\end{table*}

\begin{table*}
 \caption{Readme file of the catalog with individual measurements}\label{readmemeas}
\centering
\begin{tabular} {llcl} 
\textbf{Column} & \textbf{Label} & \textbf{Units}& \textbf{Explanation}\\ \hline
1  & RAdeg & deg & Right ascension J2000.0\\ \hline 
2  & e$\_$RAdeg & arcsec & Internal error of RAdeg\\ \hline 
3  & DEdeg & deg & Declination J2000.0\\ \hline 
4  & e$\_$DEdeg & arcsec & Internal error of DEdeg \\ \hline 
5  & Vmag & mag & Magnitude transformed to the standard Johnson V magnitude\\ \hline 
6  & e$\_$Vmag & mag & Error of Vmag \\\hline 
7  & (b-y) & mag & Str\"{o}mgren (b-y) color index\\ \hline 
8  & e$\_$(b-y) & mag & Error of (b-y) \\\hline 
9  & c1 & mag & Str\"{o}mgren $c_1$ index\\ \hline 
10 & e$\_$c1 & mag & Error of c1 \\ \hline 
11 & (v-b) & mag & Str\"{o}mgren (v-b) color index\\ \hline 
12 & e$\_$(v-b) & mag & Error of (v-b)\\ \hline 
13 & m1 & mag & Str\"{o}mgren $m_1$ index\\ \hline 
14 & e$\_$m1 & mag & Error of m1 \\ \hline 
15 & Hbeta & mag & Str\"{o}mgren H$\beta$ index\\ \hline 
16 & e$\_$Hbeta & mag & Error of Hbeta\\ \hline 
17 & RAdegu & deg & Right ascension J2000.0 in the $u$ filter CCD image\\ \hline 
18 & DEdegu & deg & Declination J2000.0 in the $u$ filter CCD image\\ \hline 
19 & umag & mag & Instrumental $u$ magnitude\\ \hline 
20 & e$\_$umag & mag & Error of umag\\ \hline 
21 & umagEC & mag & Instrumental $u$ magnitude extinction corrected\\ \hline 
22 & e$\_$umagEC & mag & Error of umagEC\\ \hline 
23 & AMu & -- & Air mass for umag\\ \hline 
24 & Xu & pixels & X pixel position for umag\\ \hline 
25 & Yu & pixels & Y pixel position for umag\\ \hline 
26 & texpu & sec & Exposure time for umag\\ \hline
27 & radu & pixels & Radius used to derive umag from aperture photometry\\ \hline
28 & skyu & counts & Sky counts associated to umag\\ \hline 
29 & sdskyu & counts & Standard deviation on skyu\\ \hline 
30 & JDu & days & Julian Date for umag\\ \hline 
31 & umagco & mag & Raw instrumental $u$ magnitude from \textit{IRAF daofind}\\ \hline 
32 & idmagu & -- & ID number in the $u$ magnitude file\\ \hline    
33-48 & & & Same as Columns 17-32 for $v$ magnitude \\ \hline
49-64 & & & Same as Columns 17-32 for $b$ magnitude \\ \hline
65-80 & & & Same as Columns 17-32 for $y$ magnitude \\ \hline
81-96 & & & Same as Columns 17-32 for $H\beta_w$ magnitude \\ \hline
97-112 & & & Same as Columns 17-32 for $H\beta_n$ magnitude\\ \hline
113 & file & -- & Name of the file where the photometry comes from\\
&&&(contains information on chip number, field number, and pointing)\\ \hline  
114 & ichip & -- & WFC chip number \\ \hline  
115 & flagFA & -- & Flag indicating which of the six filters ($u,v,b,y,H\beta_w, H\beta_n$) are available\\ \hline  
116 & ID & -- & Identifier of the star which the measure belongs to\\\hline  
117 & XmN & -- & Number of measurements with the same ID\\
\end{tabular}
\end{table*}

\begin{table*}
\caption{Readme file for the catalog with mean measurements}\label{readme}
\centering
\begin{tabular} {llcl}
\textbf{Column} & \textbf{Label} & \textbf{Units} & \textbf{Description} \\\hline
1 & ID       & &ID number\\ \hline
2 & RAdeg     & deg  & Right ascension J2000.0 \\ \hline
3 & e$\_$RAdeg   & arcsec  & Internal mean error of RAdeg  \\  \hline
4 & o$\_$RAdeg  & --     & Number of measurements for RAdeg \\ \hline
5 & DEdeg      & deg & Declination J2000.0 \\ \hline
6 & e$\_$DEdeg    & arcsec & Internal mean error of DEdeg\\ \hline
7 & o$\_$DEdeg  & --    & Number of measurements for DEdeg\\\hline
8 & Vmag       & mag & Mean magnitude transformed to the standard Johnson V magnitude \\ \hline
9 & e$\_$Vmag  & mag    & Error of Vmag (error of the mean for o$\_$Vmag$\geq$2 and internal standard deviation for o$\_$Vmag$=$1)\\ \hline
10 & o$\_$Vmag  & --    & Number of measurements for Vmag \\\hline
11 & (b-y)    & mag   & Mean Str\"{o}mgren (b-y) color index \\ \hline
12 & e$\_$(b-y)& mag     & Error of (b-y) \\ \hline
13 & o$\_$(b-y) & --       & Number of measurements for (b-y) \\\hline
14 & c1  & mag     & Mean Str\"{o}mgren $c_1$ index \\ \hline
15 & e$\_$c1 & mag    & Error of c1 \\ \hline
16 & o$\_$c1 & -- & Number of measurements for c1 \\ \hline
17 & (v-b) & mag      & Mean Str\"{o}mgren (v-b) color index  \\ \hline
18 & e$\_$(v-b)& mag     & Error of (v-b) \\ \hline
19 & o$\_$(v-b) & --       & Number of measurements for (v-b)\\\hline
20 & m1 & mag     & Mean Str\"{o}mgren $m_1$ index \\ \hline
21  & e$\_$m1& mag     & Error of m1\\ \hline
22 & o$\_$m1  & --      & Number of measurements for m1 \\ \hline
23 & Hbeta & mag      & Mean Str\"{o}mgren H$\beta$ index \\ \hline
24 & e$\_$Hbeta& mag     & Error of Hbeta \\ \hline
25 & o$\_$Hbeta  & --      & Number of measurements for Hbeta\\\hline
26 & meanJD  & days   & Mean Julian Date \\ \hline
27 & flagIA & -- & Flag indicating which of the six indexes ($V, (b-y), c1, m1, (v-b), H\beta$) are available\\ \hline
28 & flagTS & -- & Flag indicating how many measures for each index are inconsistent according to a Student's t-test\\ \hline 
29 & GSC2   & --    & GSC2 identifier\\ \hline
\end{tabular}
\end{table*}

\begin{table*}
\caption{Example of the first ten rows of the catalog with mean measurements (see Table \ref{readme})}\label{cat}
\centering
\begin{tabular} {cccccccccc}\hline
ID &  RAdeg 	& e$\_$RAdeg & o$\_$RAdeg & DEdeg & e$\_$DEdeg &
o$\_$DEdeg & Vmag &  e$\_$Vmag & o$\_$Vmag \\ \hline

     1& 83.7644656&  0.025 &  4& 30.1846810 &  0.018 &  2 &   11.836  &  0.048 &  4\\ 
     2& 83.7602742&  0.020 &  3& 30.0734584 &  0.011 &  1 &   14.088  &  0.328 &  2\\ 
     3& 83.7323671&  0.010 &  3& 30.2406015 &  0.019 &  4 &   14.081  &  0.046 &  4\\ 
     4& 83.7099509&  0.162 &  4& 30.1193681 &  0.131 &  3 &   14.325  &  0.056 &  3\\ 
     5& 83.7027602&  0.140 &  4& 30.1502936 &  0.001 &  2 &   13.543  &  0.045 &  4\\ 
     6& 83.7020158&  0.002 &  2& 30.1233125 &  0.061 &  2 &   14.272  &  0.025 &  1\\ 
     7& 83.6966936&  0.243 &  2& 30.1856348 &  0.018 &  2 &   14.797  &  0.026 &  1\\ 
     8& 83.6437971&  0.175 &  3& 30.1128599 &  0.001 &  2 &   12.742  &  0.078 &  3\\ 
     9& 83.6371473&  0.029 &  3& 30.0746250 &  0.053 &  3 &   13.914  &  0.041 &  3\\ 
    10& 83.6125620&  0.032 &  3& 30.0857448 &  0.014 &  3 &   13.706  &  0.064 &  3\\ \hline
 & (b-y) & e$\_$(b-y) & o$\_$(b-y) & c1 & e$\_$c1 & o$\_$c1 & (v-b) & e$\_$(v-b) & o$\_$(v-b) \\ \hline
&   0.327  &   0.177 &  4   &  1.014&    0.022 &  3  &   0.446 &    0.006  & 3  \\    
&   0.455  &   0.354 &  2   &  1.043&    0.043 &  2  &   0.643 &    0.187  & 2  \\    
&   0.495  &   0.007 &  4   &  0.638&    0.023 &  3  &   0.695 &    0.008  & 3  \\    
&   0.502  &   0.008 &  3   &  0.813&    0.003 &  3  &   0.597 &    0.009  & 3  \\    
&   0.938  &   0.010 &  4   &  0.376&    0.022 &  4  &   1.073 &    0.012  & 4  \\    
&   0.750  &   0.050 &  1   &  0.316&    0.075 &  1  &   0.933 &    0.057  & 1  \\    
&   0.387  &   0.053 &  1   &  1.009&    0.083 &  1  &   0.541 &    0.059  & 1  \\    
&   1.019  &   0.014 &  3   &  0.468&    0.041 &  3  &   1.232 &    0.023  & 3  \\    
&   0.662  &   0.012 &  3   &  0.453&    0.014 &  3  &   0.710 &    0.015  & 3  \\    
&   0.562  &   0.010 &  3   &  0.962&    0.009 &  3  &   0.575 &    0.007  & 3  \\ \hline
    & m1 & e$\_$m1 & o$\_$m1 & Hbeta & e$\_$Hbeta & o$\_$Hbeta & && \\ \hline
& 0.044  & 0.397 & 4 & 2.924 & 0.018 & 4 &  && \\ 
& 0.166  & 0.574 & 2 & 2.975 & 0.055 & 4 &  && \\ 
& 0.200  & 0.015 & 3 & 2.806 & 0.008 & 4 &  && \\ 
& 0.095  & 0.017 & 3 & 2.851 & 0.010 & 3 &  && \\ 
& 0.134  & 0.021 & 4 & 2.640 & 0.000 & 2 &  && \\ 
& 0.183  & 0.076 & 1 & 2.630 & 0.017 & 1 &  && \\ 
& 0.154  & 0.079 & 1 & 2.979 & 0.034 & 1 &  && \\ 
& 0.213  & 0.037 & 3 & 2.647 & 0.003 & 2 &  && \\ 
& 0.049  & 0.027 & 3 & 2.682 & 0.003 & 3 &  && \\ 
& 0.013  & 0.017 & 3 & 2.832 & 0.012 & 3 &  && \\ 
  &\multicolumn{3}{c}{meanJD} &  flagIA & flagTS &\multicolumn{2}{c}{GSC2} & &\\ \hline
&  \multicolumn{3}{c}{2455582.5239636} & 111111 & 00010010 & \multicolumn{2}{c}{N9UG000329} & &\\
&  \multicolumn{3}{c}{2455582.5239636} & 111111 & 00110110 & \multicolumn{2}{c}{N9UG000390} & &\\
&  \multicolumn{3}{c}{2455582.5239636} & 111111 & 00000000 & \multicolumn{2}{c}{N9UG000301} & &\\
&  \multicolumn{3}{c}{2455582.5239636} & 111111 & 00000000 & \multicolumn{2}{c}{N9UG000367} & &\\
&  \multicolumn{3}{c}{2455582.5239636} & 111111 & 10000000 & \multicolumn{2}{c}{N9UG000351} & &\\
&  \multicolumn{3}{c}{2455591.5528230} & 111111 & 01000000 & \multicolumn{2}{c}{N9UG000365} & &\\
&  \multicolumn{3}{c}{2455591.5536240} & 111111 & 10000000 & \multicolumn{2}{c}{N9UG015193} & &\\
&  \multicolumn{3}{c}{2455585.5365401} & 111111 & 00000000 & \multicolumn{2}{c}{N9UG000370} & &\\
&  \multicolumn{3}{c}{2455585.5365401} & 111111 & 00000000 & \multicolumn{2}{c}{N9UG000387} & &\\
&  \multicolumn{3}{c}{2455585.5365401} & 111111 & 00000000 & \multicolumn{2}{c}{N9UG000383} & &\\ \hline
\end{tabular}
\end{table*}

\end{document}